# Whispering-gallery-mode based $CH_3NH_3PbBr_3$ perovskite microrod lasers with high quality factors

Kaiyang Wang[a, #], Shang Sun[b, #], Chen Zhang[b], Wenzhao Sun[a], Zhiyuan Gu[a], Shumin Xiao[b, *], Qinghai Song[a, c, †]

a. Integrated Nanoscience Lab, Department of Electrical and Information Engineering, Harbin Institute of Technology, Shenzhen, China, 518055.

b. Integrated Nanoscience Lab, Department of Material Science and Engineering, Harbin Institute of Technology, Shenzhen, China, 518055.

c. State Key Laboratory on Tunable laser Technology, Harbin Institute of Technology, Harbin, China, 158001.

d. * Email: shumin.xiao@hitsz.edu.cn

e. † Email: qinghai.song@hitsz.edu.cn

f. # These two authors contribute equally to this manuscript.

Lead halide perovskite based micro- and nano- lasers have been widely studied in past two years. Due to their long carrier diffusion length and high external quantum efficiency, lead halide perovskites have been considered to have bright future in optoelectronic devices, especially in the "green gap" wavelength region. However, the quality (Q) factors of perovskite lasers are unspectacular compared to conventional microdisk lasers. The record value of full width at half maximum (FWHM) at threshold is still around 0.22 nm. Herein we synthesized solution-processed, single-crystalline CH3NH3PbBr3 perovskite microrods and studied their lasing actions. In contrast to entirely pumping a microrod on substrate, we partially excited the microrods that were hanging in the air. Consequently, single-mode or few-mode laser emissions have been successfully obtained from the whispering-gallery like diamond modes, which are confined by total internal reflection within the transverse plane. Owning to the better light confinement and high crystal quality, the FWHM at threshold have been significantly improved. The smallest FWHM at threshold is around 0.1 nm, giving a Q factor over 5000.

**Introduction**

Due to their ultralow defect density and long carrier diffusion length, lead halide perovskites ($CH_3NH_3PbX_3$) have been demonstrated as efficient photovoltaic materials [1]. In just four years, the light conversion coefficient has been dramatically increased from around 3% to over 20% [2-7]. Very recently, these unique properties of lead halide perovskites have been utilized in optical and photonic devices [8, 9]. Light emitting devices, especially the micro- & nano- lasers, have been successfully demonstrated in both polycrystalline films [8, 9] and single-crystal microstructures [10-17]. In additional to the solution-processed synthesis, lead halide perovskite optoelectronic devices have shown their advantages in high external quantum efficiency [8-10] and low carrier density at threshold. Most importantly, their emission wavelengths can be controlled from ultraviolet to near infrared by changing the halide from Cl, Br, I [8]. This suddenly opens important applications of $CH_3NH_3PbBr_3$ perovskite [8, 11-14] in green-light regions (540 nm – 580 nm, also known as "green gap"), where the conventional semiconductors such as III-N and III-P face severe challenge. To be commercially available, the long-term stability of the lead halide perovskite needs to be significantly improved. Here it should be noted that recently formamidinium lead halide perovskite (FAPbX$_3$) microlasers, which feature red-shifted emission and better photo and thermal stability compared to MAPbX$_3$, has been demonstrated [18].

While the lead-halide perovskite lasers have been widely studied in polygon microcavities [14-17], microrods [11-13], nanorods [10], spherical cavities [19], as well as random lasers in microcrystal network or clusters [20, 21], their quality (Q) factors are usually much lower than conventional microlasers. The reported full width at half maximum (FWHM) at lasing threshold in hexagonal microdisks and square microdisks are usually larger than 0.5 nm [14-17]. In 2015, Zhu et al. have improved the crystal quality of perovskite and reduced the FWHM at threshold to record low values around 0.22 nm [10]. However, this value is still far from the one of GaAs based circular microdisk or deformed microdisk [22]. Herein, we synthesize the CH$_3$NH$_3$PbBr$_3$ perovskite microrods and study their lasing actions. In general, the quality factor is dependent on the absorption loss, radiation loss, and the scattering loss following the equation $1/Q = 1/Q\_rad + 1/Q\_sca + 1/Q\_abs$. The absorption loss and scattering loss can be decreased by improving the crystal quality, whereas the radiation loss is mostly associated with the resonators and some additional mechanisms. By improving both the crystal quality and the cavity design, we have further decreased the FWHM at lasing threshold to around 0.1 nm, giving a Q factor over 5000.

**Results and discussion**

The microrods were synthesized with the solution-processed one-step precipitation method. CH$_3$NH$_3$Br and PbBr$_2$ were independently dissolved in N,N-dimethylformamide (DMF) with concentrations around 0.1 M [12]. Then two solutions were mixed with 1:1 volume ratio to form CH3NH3Br·PbBr$_2$ solution (0.05 M). The mixed solution was dip-casted onto an ITO-coated glass substrate, which was placed on a teflon stage in a 250 ml beaker. 75 ml dichloromethane (DCM) of CH$_2$Cl$_2$ was added in the beaker, which was sealed with a porous Parafilm (3 M). After 24 hours, CH$_3$NH$_3$PbBr$_3$ perovskites have been successfully synthesized on the substrate. Figure 1(a) shows the top-view scanning electron microscope (SEM) image of the synthesized CH$_3$NH$_3$PbBr$_3$ perovskites. Within the measured area, we can see that the cuboid microrods are dominant structures. The lengths and widths of microrods, which can be directly taken from the SEM images, vary in the range from a few micrometers to tens of micrometers and a few hundred nanometers to a few microns, respectively. The naturally grown microrods usually have square cross sections (see Figure 2(a) as examples). The insets in Figures 4(b), 4(d), 4(f) show the high resolution SEM images of microrods. We can see that the facets of microrods are extremely flat and only a few flaws randomly distribute on them. Figure 1(b) shows the X-ray diffraction result of the perovskite microrods. A set of diffraction peaks at 2θ = 14.7° (001) and 29.86° (002) are consistent with the Zhu's report well [10] and clearly shows the cubic perovskite phase. All these information shows the high quality of synthesized perovskites and indicates the possibility of achieving high Q laser.

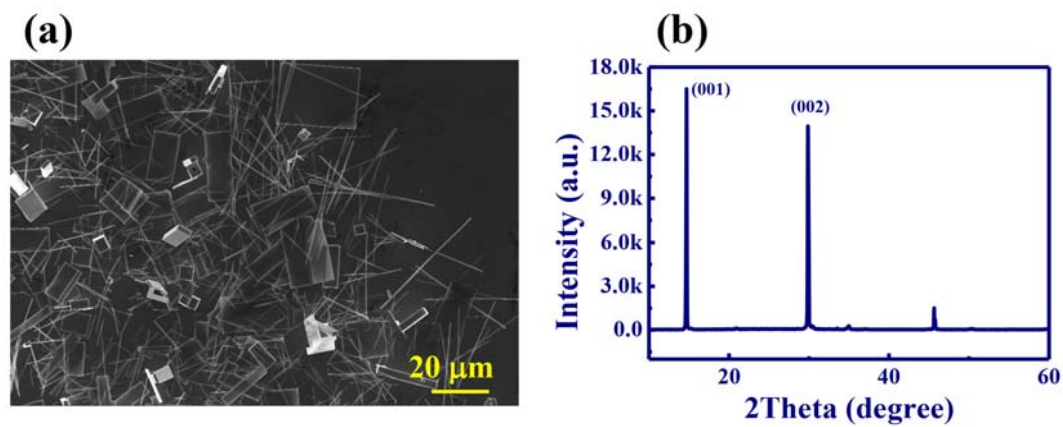

Fig. 1 The SEM image (a) and XRD (b) of synthesized perovskite microrods.

Then the optical properties of the synthesized microrods were studied by optically exciting them under a home-made micro-photoluminescence system (see experimental section). We first tested the conventional Fabry-Perot (F-P) modes along one microrod on the substrate. With the increase of pumping power, sharp peaks emerged in the laser spectrum. However, as what we have reported recently, these F-

P lasers usually have very large FWHM and very low factors (~ a few hundred) [12]. This kind of low Q factors can be simply understood. The F-P mode along the axis of microrod is usually confined by the reflections between two end-facets. While the refractive index of lead halide perovskite is relatively high (n ~ 2.55), the transmission losses are still large. In this sense, the Q factors of F-P modes along the axis of microrod are usually low and they are unable to support high Q lasers.

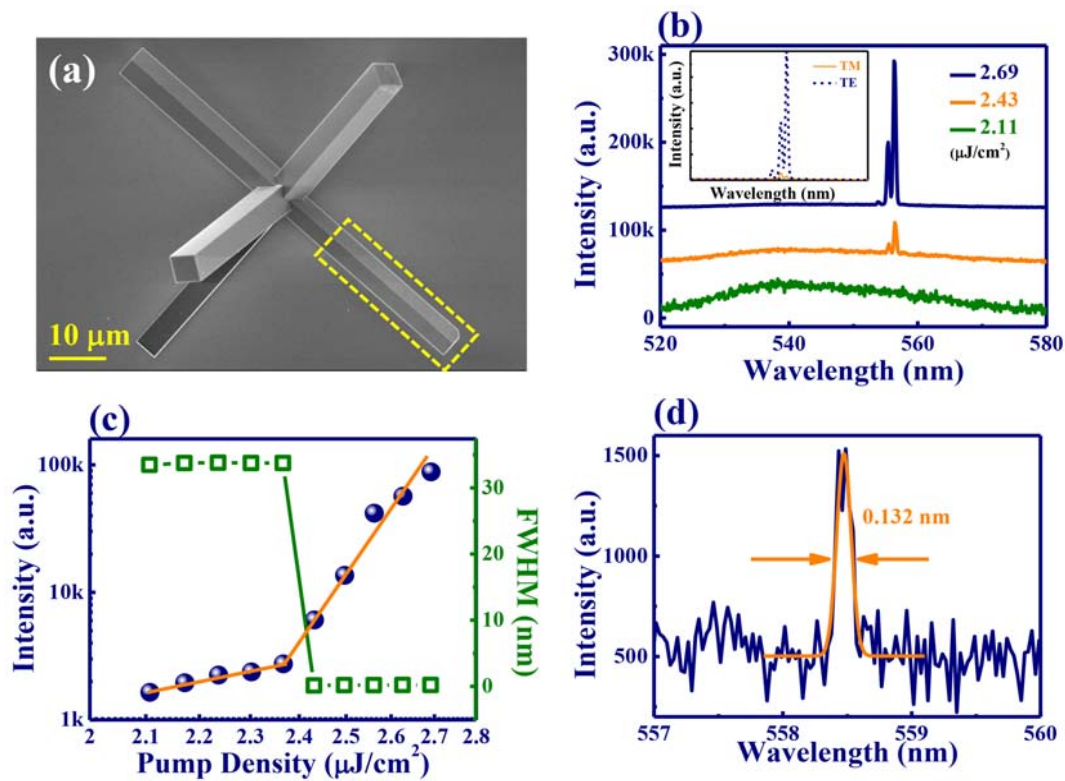

**Fig. 2** The transverse lasing action. (a) SEM image of the big "Cross". The area under the yellow dashed rectangular box is the optical pumping area. (b) The emission spectra from selected perovskite microrod under different pumping density. The inset shows the polarization of emitted light. (c) The output intensity (spheres) and FWHM (squares) as a function of pumping density. (d) The laser spectra and fitted FWHM at threshold.

In additional to F-P modes, the cross-sections of perovskite microrods are large enough to support another type of resonances in the transverse plane. In general microrod or nanorod, such kind of resonances cannot be excited because F-P modes have much longer amplification lengths. In some particular designs, the F-P modes can be suppressed and the transverse lasing modes can be excited. One example is shown in Figure 2. From its SEM image in Figure 2(a), we can see that several microrods joint together and form a big cross. As the reflections of waveguide modes between two end-facets will be broken at the joint position, the radiation loss of F-P modes along the arms of the cross increases significantly and the Q factors of F-P modes are strongly spoiled. Figure 2(b) shows the recorded laser spectra of the cross by partially pumping the selected microrod (see Figure 2(a)). When the pumping density is low, a broad photoluminescence peak has been achieved. With the increase of pumping power, a few narrow peaks appeared around 556.4 nm and quickly dominated the emission spectra. One additional laser peak appeared with further increase in pumping power. The spectra of light emissions from partially pumped microrod are quite different from the reported F-P microrod lasers and are consistent with above analysis well.

The emergence of narrow peaks usually corresponds to the transition from spontaneous emission to lasing actions. And this kind of transition can be seen more clearly in Figure 2(c). When the pumping density is below 2.37 $\mu J/cm^2$, the slope of output intensity is around 0.8. Once the pumping density is above 2.37 $\mu J/cm^2$, the slope dramatically increases to ~ 8. The superliner slope and the sharp peaks clearly demonstrate a threshold behavior at 2.37 $\mu J/cm^2$. The squares in Figure 2(c) illustrate the laser FWHM as a function of pumping density. When the pumping density is below threshold, the FWHM is over 30 nm, which is consistent with the spontaneous emission well. Once the pumping density is above threshold, the

FWHM is dramatically reduced by more than two orders of magnitude. Due to the band filling effect [23], the FWHM increased again above the threshold. Thus the smallest FWHM at threshold, which is as small as 0.132 nm, is achieved at laser threshold. This value is almost half of the previous reported record narrow laser peak and gives a high Q factor around 4200. The FWHM and Q factor can be even better by getting rid of the instrumental broadening of our spectrometer (~ 0.11 nm).

To here we know a surprising fact that the FWHM can be simply improved by exciting the transverse lasing modes. To understand the experimental observations, we have numerically studied the microrods with finite element method based commercial software (Comsol Multiphysics 3.5a) [24]. All the parameters of perovskite microrod are taken from SEM images. Considering the situation of microrod, a two-dimensional object has been employed to mimic the lasing properties. The width and thickness of the selected microrod under consideration are both 1.964 μm. The refractive index of perovskite is fixed at 2.55. As the microrod is standing on one edge, the surrounding medium is set as n = 1. Following the experimental results, only transverse electric (TE) modes are considered. Figure 3(a) shows the calculated Q factors (Q= Re(ω)/2|Im(ω)|, here ω is the resonant frequency) around the gain spectral region. We can see that most of the resonances have relatively low Q factors except modes 1-3 marked in Figure 3(a). This is because that the resonances inside rectangle shaped microcavities follow the equation [24]

$$\psi(x,y) = \sin\left(\frac{m\pi}{a}x\right)\sin\left(\frac{l\pi}{b}y\right), \qquad (1)$$

if 0≤x≤a and 0≤y≤b. Here a, b are the lengths of two sides of rectangle and m, l are the numbers of maximal |ψ(x,y)|². From Eq. (1), it is easy to see that the resonances in rectangle cavities should follow the same behaviors as F-P modes and thus have quite low Q factors around a few hundred. Interestingly, due to the two-dimensional confinements, some resonances in x-direction can have similar frequencies to the resonances in y-direction. Consequently, they can interact with each other and some whispering-gallery like modes have been formed (see Figure 3(b)) [24]. Because their field distributions of such kind modes are destructively cancelled around the corners, the main leakages around the corners are suppressed and high Q factors have been formed [25-31]. As shown in Figure 3(a), the Q factors of modes 1-3 are almost two orders of magnitude larger. Therefore, the FWHM of laser in transverse plane won't be limited by the cavity radiation loss anymore and the performances of perovskite laser can be improved.

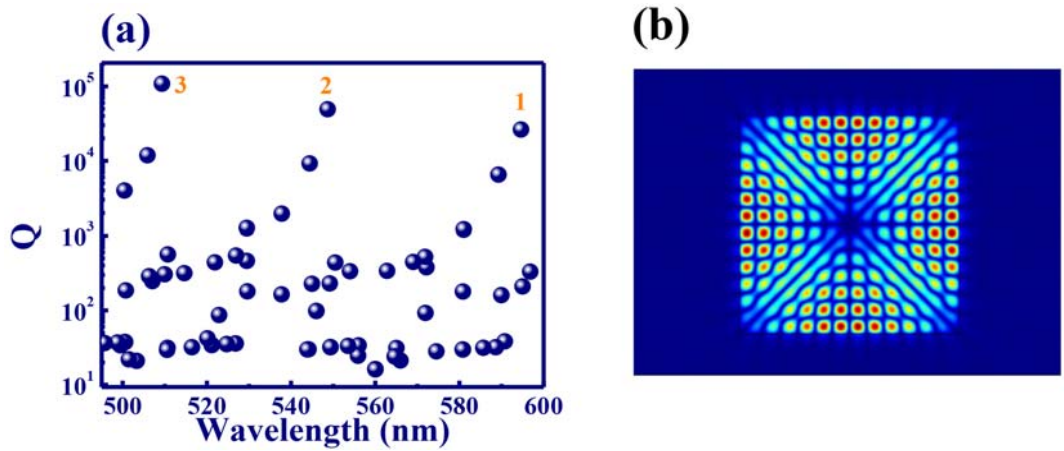

**Fig. 3** Resonant modes in transverse plane of microrod. (a) The calculated Q factors of resonances inside the transverse plane of the microrod marked in Figure 2(a). (b) The field distribution (|Hz|) of mode-2 in (a).

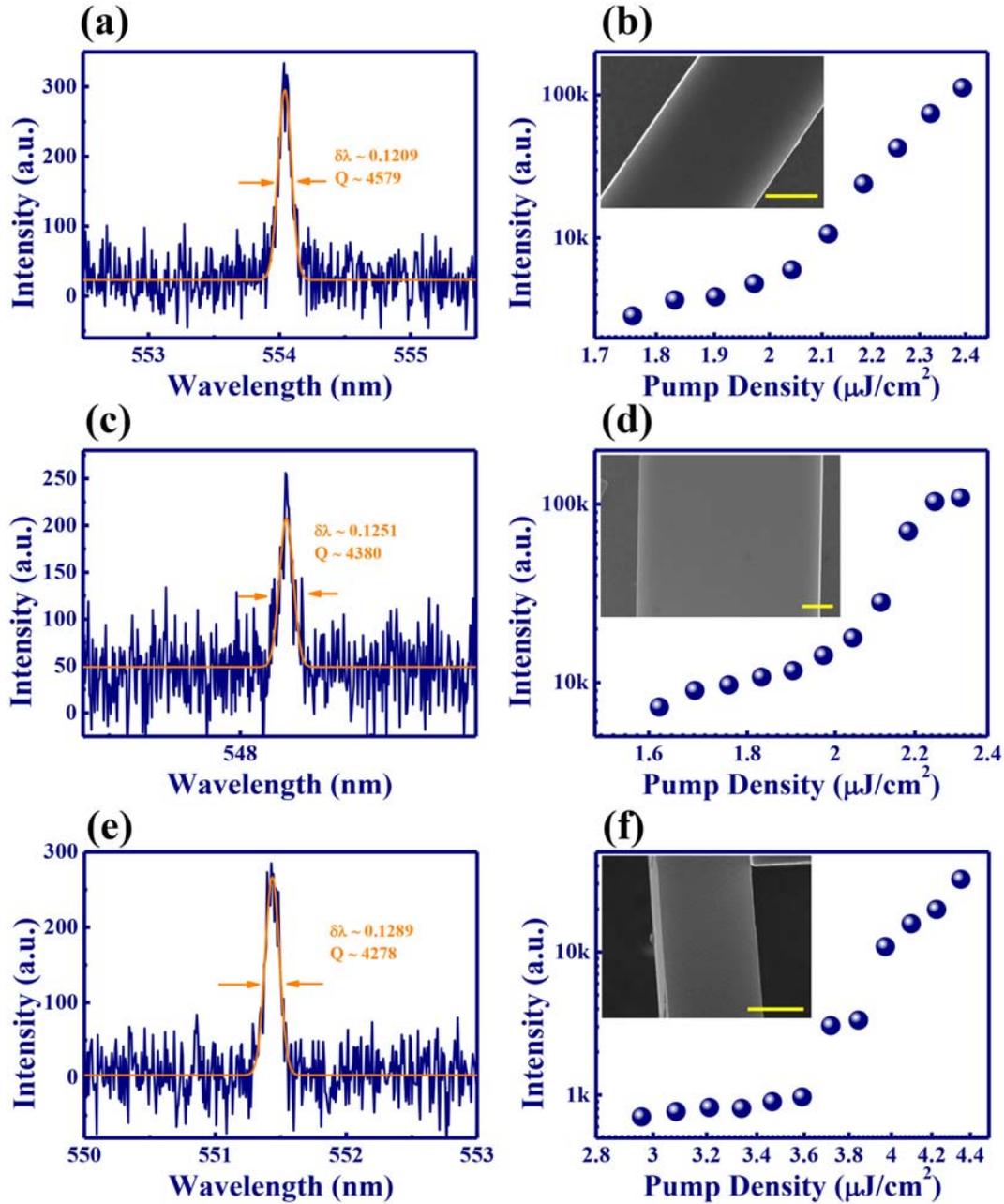

**Fig. 4** FWHMs in different microrods. (a), (c), and (e) show the laser spectra above threshold. And (b), (d), (f) show the dependences of output intensities on the pumping densities. The insets are the high resolution SEM images of the measured perovskite microrods and the scale bar is 1 μm.

We note that the high Q factors are not limited in the particular sample. Similar phenomena have been widely observed in a number of microrods. Parts of the results are shown in Figure 4. Here we excited the transverse lasing modes of microrods suspending in air. The samples could be found from randomly synthesized perovskite microrods or constructed by placing the microrods on a perovskite blocks with one end suspending in air via micro-manipulation [12]. Because the perovskite block can introduce significant radiation loss in the longitudinal direction, the F-P lasing modes can also be suppressed and transverse modes are excited. From the insets in Figures 4(b), 4(d), and 4(f), we know that these microrods are also quite uniform and flat, providing the chances to reach small FWHM. The experimental results are shown in Figure 4. While their exact transvers shapes and thresholds are different, their FWHMs at lasing threshold are always around 0.1 nm. The details of microrods sizes, measured and numerically calculated Q factors are listed in Table 1. It should be noted that the lasing spectra shown in Figure 4 were taken above threshold to maintain a high signal to noise ratio (SNR). Considering the band filling effect [23], the FWHM at threshold is the smallest. To obtain the exact and true FWHMs, the FWHMs (corresponding to Q factors) in Table 1 are extracted at lasing threshold. We know that the smallest FWHM can be as small as 0.1 nm, giving a Q factor over 5000. Interestingly, we found that almost all the calculated Q factors were much larger than the experimentally measured values. This means that the current Q factors are limited by the material instead of cavity shape again and thus they can be further increased by improving the crystal quality.

One may also wonder why the high Q factors have not been observed in the microplate lasers, whose resonances are also confined by total internal reflections [14-17]. In principle, the microplates have much larger cavity sizes and should have much larger simulated Q factors. But the experimentally recorded FWHMs are usually around or far larger than 0.5 nm [14-17]. Some values are almost an order of magnitude larger than our observations [14-17]. This deviation is generated by the substrate. Different from the microrod in Figure 2(a), all the microplates are attached on the substrate after the synthesis. Thus the influences of substrates must be considered. Compared with the two-dimensional calculation, there are vertical losses into the substrate. Sometimes, the vertical losses are even much larger. For example, when the microrod in Figure 3 is placed onto a glass substrate, the Q factor of mode-2 will be degraded by more than an order of magnitude. Once the ITO (n = 1.8) layer on the substrate is considered, the Q factor is even lower. Meanwhile, most substrates cannot be as flat as the single-crystalline samples. Additional scattering losses will also decrease the cavity Q factors. In addition, there are always some flaws in synthesized devices (see insets in Figures 4(a), 4(c), and 4(e)). The transverse modes can be selectively formed at the flawless positions. However, the resonances in microplates and F-P modes in microrods cover most of the devices and cannot fully avoid such scattering centers. Therefore, long-lived resonances are much easier to be formed in the transverse planes of freestanding microrods than F-P modes in microrods and whispering gallery like modes in microplates.

**Table 1** The measured Q factors of transverse lasers in different microrods. The cavity sizes and their simulated Q factors are also listed.

| Sample# | Width ($\mu$m) | Height ($\mu$m) | Q_mea | Q_sim |
| --- | --- | --- | --- | --- |
| 1 | 1.469 | 1.394 | 5968 | 8763 |
| 2 | 1.537 | 1.456 | 5697 | 5175 |
| 3 | 2.573 | 1.897 | 4971 | 27506 |
| 4 | 2.516 | 2.516 | 4846 | 257561 |
| 5 | 1.491 | 1.491 | 4417 | 15744 |
| 6 | 0.895 | 0.895 | 4384 | 4205 |
| 7 | 5.969 | 0.7 | 4380 | 3332 |
| 8 | 1.303 | 1.303 | 4302 | 8973 |
| 9 | 2.886 | 2.886 | 4215 | 330205 |
| 10 | 1.401 | 1.401 | 4207 | 26967 |

## Conclusions

In summary, we have studied the whispering-gallery like modes in the transverse plane of freestanding microrod. The smallest FWHM at lasing threshold is 0.1 nm, which is several times smaller than the previously reported record values in lead halide perovskite nanorod. And the measured FWHM can even be comparable with GaAs microdisk lasers. Compared with the microplates, perovskite microrods have additional advantages. They can be easily tailored by micro-manipulation to improve the laser performances. In addition, the sizes of perovskite microrods are usually quite uniform. The emitted laser peaks from different positions on the same microrod are almost identical or quite similar. Thus the microrod can be either pumped entirely to achieve high output power or pumped individually to achieve parallel laser array [32]. We believe our finding will be essential for the practical applications of lead-halide perovskite microlasers.

## Experimental Section

The optically pumped lasing experiment was performed on a home-made micro-photoluminescence system. Basically, the samples were placed onto a three-dimensional translation stage and a frequency doubled Ti:Sapphire laser from regenerative amplifier (400 nm, 1k Hz, 100 fs pulse duration) was focused by a 40× objective lens onto the top-surface of the samples. The emitted light was collected by the same objective lens and coupled to a spectrometer via a multimode fiber. A polarizer was placed in front of the fiber to record the polarization of laser emission.

## Acknowledgements

The author would like to thank the financial support from National Natural Science Foundation of China under the Grant No. 11374078; Shenzhen Peacock plan (KQCX20130627094615410); and Shenzhen Fundamental research projects (JCYJ20140417172417110, JCYJ20140417172417096).

## References

1  M. A. Green, A. Ho-Baillie, and H. J. Snaith, *Nature Photonics,* 2014, **8**, 506-514.
2  A. Kojima, K. Teshima, Y. Shirai and T. Miyasaka, *J. Am. Chem. Soc.,* 2009, **131**, 6050–6051.
3  M. M. Lee, J. Teuscher, T. Miyasaka, T. N. Murakami and H. J. Snaith, *Science,* 2012*,* **338,** 643–647.
4  H.-S. Kim, C.-R. Lee, J.-H. Im, K.-B. Lee, T. Moehl, A. Marchioro, S.-J. Moon, R. Humphry-Baker, J.-H Yum, J. E. Moser, M. Gra¨tzel and N.-G. Park, *Sci. Rep.,* 2012, **2,** 591.
5  N. J. Jeon, J. H. Noh, W. S. Yang, Y. C. Kim, S.-C. Ryu, J.-W. Seo and S. I. Seok*, Nature,* 2015, **517,** 476–480.
6  National Renewable Energy Labs (NREL) efficiency chart (2015); http://www.nrel.gov/ncpv/images/efficiency_chart.jpg.
7  W. S. Yang, J. H. Noh, N. J. Jeon, Y. C. Kim, S.-C. Ryu, J.-W. Seo and S. I. Seok, *Science*, 2015, **348**, 1234-1237.
8  G. C. Xing, N. Mathews, S. S. Lim, N. Yantara, X. F. Liu, D. Sabba, M. Grätzel, S. Mhaisalkar and T. C. Sum, *Nat. Mater.*, 2014, **13**, 476-480.
9  F. Deschler, M. Price, S. Pathak, L. E. Klintberg, D. D. Jarausch, R. Higler, S. Hüttner, T. Leijtens, S. D. Stranks, H. J. Snaith, M. Atatüre, R. T. Phillips, and R. H. Friend, *J. Phys. Chem. Lett.*, 2014, **5,** 1421–1426.
10 H. M. Zhu, Y. P. Fu, F. Meng, X. X. Wu, Z. Z. Gong, Q. Ding, M. V. Gustafsson,M. T. Trinh, S. Jin2 and X. Y. Zhu, *Nat. Mater.* 2015, **14**, 636-642.
11 Z. Y. Gu, K. Y. Wang, W. Z. Sun, J. K. Li, S. Liu, Q. H. Song, H. Cao, *Adv. Opt. Mat.,* 2016, **4**, 472-479.
12 K. Y. Wang, Z. Y. Gu, S. Liu, J. K. Li, S. M. Xiao, Q. H. Song, *Opt. Lett.* 2016, **41**, 555-558.
13 W. Z. Sun, K. Y. Wang, Z. Y. Gu, S. M. Xiao and Q. H. Song, *Nanoscale*, 2016, DOI: 10.1039/C6NR00436A.
14 Q. Liao , K. Hu , H. H. Zhang , X. D. Wang , J. N. Yao , and H. B. Fu, *Adv. Mater.* 2015, **27**, 3405–3410.
15 Q. Zhang, S. T. Ha, X. F. Liu,T. C. Sum and Q. H. Xiong, *Nano Lett.*, 2014, **14**, 5995-6001.
16 S. T. Ha , X. F. Liu , Q. Zhang , D. Giovanni , T. C. Sum and Q. H. Xiong, *Adv. Opt. Mater.* 2014, **2**, 838–844.
17 X. F. Liu, S. T. Ha, Q. Zhang, M. D. L. Mata, C. Magen, J. Arbiol, T. C. Sum and Q. H. Xiong, *ACS Nano*, 2015, **9**,687–695.
18 Y. P. Fu, H. M. Zhu, A. W. Schrader, D. Liang, Q. Ding, P. Joshi, L. Y. Hwang, X-Y. Zhu, S. Jin, *Nano Lett.*, 2016, **16**, 1000-1008.
19 B. R. Sutherland, S. Hoogland, M. M. Adachi, C. T. O. Wong, and E. H. Sargent, *ACS Nano*, 2014, **8**, 10947–10952.
20 R. Dhanker, A. N. Brigeman, A. V. Larsen, R. J. Stewart, J. B. Asbury and N. C. Giebink, *Appl. Phys. Lett.*, 2014, **105**, 151112.
21 S. Yakunin, L. Protesescu, F. Krieg, M. I. Bodnarchuk, G. Nedelcu, M. Humer, G. D. Luca, M. Fiebig, W. Heiss, M. V. Kovalenko, *Nat. Commun.*, 2015, **6**, 8056.
22 Q. Song, H. Cao, S. T. Ho and G. S. Solomon, *Appl. Phys. Lett.*, 2009, **94**, 061109.
23 B. Hua, J. Motohisa, Y. Kobayashi, S. Hara, and T. Fukui, *Nano Lett.*, 2008, **9**, 112-116.
24 S Liu, Z. Y. Gu, N. Zhang, K. Y. Wang, Q. Lyu, K. Xu, S. M. Xiao, and Q. H. Song, *J. Lightwave Technol.*, 2015, **33**, 3698-3703S
25 J. Wiersig, *Phys. Rev. A,* 2003, **67**, 023807.
26 Q. H. Song and H. Cao, *Phys. Rev. Lett.,* 2010, **105**, 053902.
27 Y. Z. Huang, K. J. Che, Y.-D. Yang, S. J. Wang, Y. Du, Z. C. Fan, *Opt. Lett.* 2008, **33**, 2170-2172.
28 Y. Z. Huang, Y. H. Hu, Q. Chen, S. J. Wang, Y. Du, Z. C. Fan, *IEEE Photon. Tech. Lett.* 2007, **19**, 963-965.
29 Q. H. Song, L. Ge, J. Wiersig and H. Cao, *Phy. Rev. A*, 2013, **88**, 023834.
30 A. W. Poon, F. Courvoisier, and R. K. Chang, Opt. Lett., 2001, **26**, 632-634.
31 M. Lebental, N. Djellali, C. Arnaud, J.-S. Lauret, J. Zyss, R. Dubertrand, C. Schmit, and E. Bogomolny, *Phys. Rev. A,* 2007, **76**, 023830.
32 Q. Liao, X. Jin, H. H. Zhang, Z. Z. Xu, J. N. Yao, and H. B. Fu, *Angew. Chem. Int. Ed.,* 2015, **54**, 7037-7041.